\documentclass[aps,twocolumn,prb,10pt,floatfix]{revtex4-2}

\setcitestyle{super,open={},close={}}
\usepackage{pdfcomment} 	
\usepackage[UKenglish]{isodate} 

\usepackage{amsmath} 		
\usepackage{amssymb} 		
\usepackage{chemscheme}     
\usepackage{chemformula}    

\usepackage{bm} 			
\usepackage{breqn} 			

\usepackage{booktabs}	    
\usepackage{multirow}	    
\usepackage[flushleft]{threeparttable}
\usepackage{floatrow}       
\usepackage{color}
	\definecolor{mygray}{rgb}{0.8, 0.8, 0.8}
	\definecolor{coolblack}{rgb}{0.0, 0.18, 0.39}
        \newcommand{\Black}[1] {{\color{black} #1}} 	
\usepackage{lipsum}         
\usepackage{hyperref}       
	\hypersetup{colorlinks=true,allcolors=coolblack}

\usepackage{epstopdf} 					
\begin{document}
\date{\today}

\author{Philipp J. Heckmeier*$^{,1}$, Jeannette Ruf$^{1}$, Charlotte Rochereau$^{2}$, and Peter Hamm$^{1}$\\\textit{$^{1}$Department of Chemistry, University of Zurich, Zurich, Switzerland}\\
\textit{$^{2}$Department of Systems Biology, Columbia University, USA}\\
$^*$philipp.heckmeier@chem.uzh.ch}

\title {A billion years of evolution manifest in nanosecond protein dynamics}

\begin{abstract}
{\textbf{Protein dynamics form a critical bridge between protein structure and function\cite{Karplus2005,Boehr2009}, yet the impact of evolution on ultrafast processes inside proteins remains enigmatic. This study delves deep into nanosecond-scale protein dynamics of a structurally and functionally conserved protein across species separated by almost a billion years \cite{Adams2007,Sora2022,Kumar2022}, investigating ten homologs in complex with their ligand. By inducing a photo-triggered destabilization of the ligand inside the binding pocket \cite{Jankovic2021, Heckmeier2022}, we resolved distinct kinetic footprints for each homolog via transient infrared spectroscopy \cite{Lorenz-Fonfria2020,Barth2007}. Strikingly, we found a cascade of rearrangements within the protein complex which manifest in three discrete time points of dynamic activity, conserved over hundreds of millions of years within a narrow window. Among these processes, one displays a subtle temporal shift correlating with evolutionary divergence, suggesting reduced selective pressure in the past. Our study not only uncovers the impact of evolution on molecular processes in a specific case, but has also the potential to initiate a novel field of scientific inquiry within molecular paleontology, where species are compared and classified based on the rapid pace of protein dynamic processes; a field which connects the shortest conceivable time scale in living matter (10$^{-9}$~s) with the largest ones (10$^{16}$~s).}}

\end{abstract}
\maketitle

\section*{Main}
Proteins exist as dynamic ensembles, rather than being rigid and static entities. They constantly undergo rearrangements, folding-, and unfolding processes on a nanosecond time scale \cite{Boehr2009,Galvanetto2023,Jankovic2021,Heckmeier2022}. Understanding this dynamic nature is essential to comprehending their function. As protein dynamics serve as the crucial link between structure and function\cite{Karplus2005}, their experimental investigation has predominantly focused on individual protein examples, providing insights into specific\cite{Schanda2005,Pirchi2011,Aviram2018}, often intrinsically disordered cases\cite{Hofmann2010,Borgia2018,Jemth2018}. Surprisingly, protein dynamics within a group of closely related proteins, such as a family of homologs, have rarely been experimentally explored, and if so, in the slow-paced millisecond-second regime\cite{Jemth2018,Karlsson2022,Schenkmayerova2021} where rapid fluctuations of conformational adaptations are not resolved. Consequently, little is known about whether structural homologs display conserved ultrafast protein dynamics throughout evolution. How may nano-scale protein dynamics evolve over hundreds of million years within a protein family?

Revealing the rapid dynamic processes within proteins requires the use of an appropriate toolkit. Thus far, the conservation of protein structures has been primarily observed through structure comparison using X-ray crystallography \cite{Ortlund2007,Ingles-Prieto2013,Nguyen2017,Hadzipasic2020}. X-ray crystallography provides valuable insights with a predominantly static view of proteins, but lacks the mechanistic intricacies that define their dynamics. As an alternative approach, NMR spectroscopy excels at resolving small conformational differences and dynamics in equilibrium \cite{Alderson2021,Jemth2018,Karlsson2022,Schanda2005}, yet it falls short in recording non-equilibrium processes. 

In contrast, infrared spectroscopy is sensitive to subtle differences in protein conformations and is a powerful tool to temporally resolve fast dynamical processes within proteins \cite{Lorenz-Fonfria2020,Barth2007}. In combination with an  phototrigger, this technique enables the initiation and monitoring of sequential destabilization within a protein complex, with a temporal resolution as fast as a picoseconds \cite{Jankovic2021,Bozovic2020b,Heckmeier2022,Helbing2023}. The key challenge lies in investigating the specific time points at which certain processes occur, in order to resolve the influence of evolution on molecules that are inherently dynamic and exhibit fluent transitions between conformational states.

\begin{figure*}[t]
	\centering
	\includegraphics[width=\textwidth]{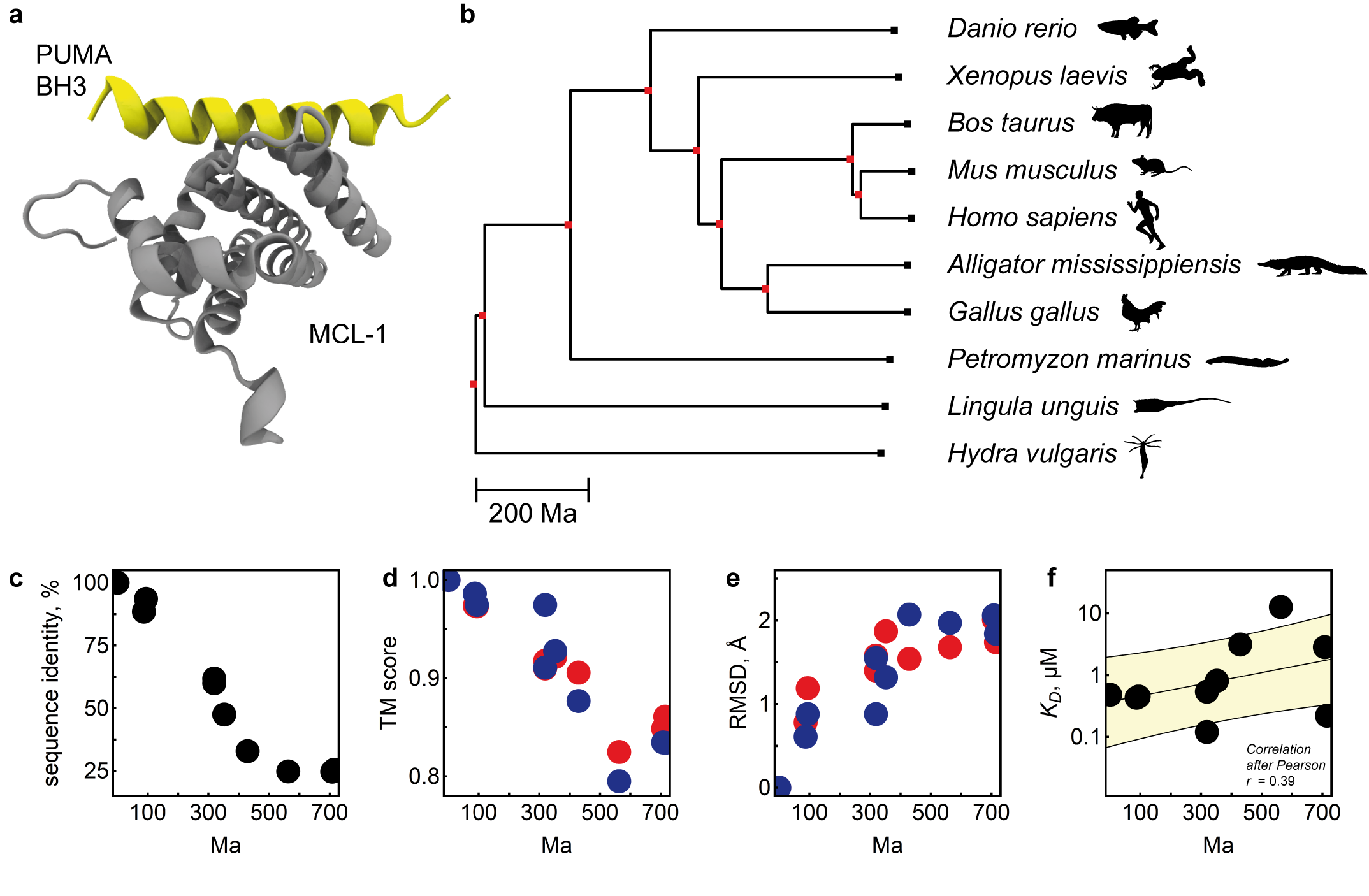}
	\caption{Structure and function of MCL-1 is conserved. (a) NMR structure of MCL-1 (grey) complexed with PUMA BH3 (yellow) (PDB: 2roc)\cite{Day2008}. (b) Phylogeny of ten species whose MCL-homologs were selected for this study. The phylogeny and the corresponding evolutionary divergence time in million years (Ma) were taken from TimeTree5 and cover the current state of science (July 2023) \cite{Kumar2022}. (c) Sequence identity of all investigated MCL-1 homologs (compared to  \textit{H. sapiens}) against evolutionary divergence time of the corresponding species. (d,e) Structural similarity between MCL-1 homologs (compared to \textit{H. sapiens}), predicted with AlphaFold\cite{Jumper2021} (blue) and RosettaFold\cite{Baek2021} (red).  (f) MCL-1 homolog binding free energy for the PUMA BH3 peptide, plotted against the evolutionary divergence time. Yellow, linear fit $\pm$ standard deviation. The Pearson correlation coefficient $r=0.39$ indicates that the binding free energy correlates weakly with evolutionary divergence time.}  \label{fig:Introduction}
\end{figure*}

\section*{MCL-1: a prime example of conservation}
This study is concerned with the  protein myeloid cell leukemia 1 (MCL-1), a member of the BCL-2 protein family, which plays a crucial role as a key regulator of apoptosis, the programmed cell death\cite{Adams2007,Youle2008}. It is found not only in humans, but also in a diverse range of metazoan organisms \cite{Aouacheria2015, Sora2022}. Functioning as an anti-apoptotic protein, MCL-1 interacts promiscuously with pro-apoptotic factors through $\alpha$-helical domains known as BCL-2 homology domain 3 (BH3)\cite{Youle2008,Kale2018,Heckmeier2023,Sora2022}, e.g. the BH3 domain  of the pro-apoptotic protein PUMA\cite{Czabotar2007, Day2008} (Fig.~\Ref{fig:Introduction}a). 
Homologs of this protein family have been identified in all vertebrates and even in more distantly related species such as sponges \cite{Wiens2001} and \textit{Cnidaria} \cite{Lasi2010}, whose last common ancestor with \textit{Homo sapiens} existed over 700 million years ago \cite{Kumar2022}. 

We selected ten MCL-1 homologs (Fig.~\Ref{fig:Introduction}b) from species, whose last common ancestors with \textit{Homo sapiens} are distributed equidistantly on an evolutionary time axis up to a billion years from present day to the past. We opted for a horizontal approach by comparing sequences of currently living species, as opposed to a vertical approach involving the reconstruction of ancestral proteins \cite{Harms2010,Merkl2016}. Besides \textit{Homo sapiens}, we included \textit{Mus musculus}, \textit{Bos taurus}, \textit{Gallus gallus}, \textit{Alligator mississippiensis}, \textit{Xenopus laevis}, \textit{Danio rerio}, a \textit{Petromyzon marinus} candidate\cite{Timoshevskaya2023}, \textit{Lingula unguis}, and \textit{Hydra vulgaris}.

Before exploring the protein dynamics for this homolog selection, our objective was to unequivocally establish the conservation of both the structure and function of MCL-1. By comparing the amino acid sequences of the homologs to their human equivalent, we found that sequence identity dramatically decreased as a function of evolutionary divergence (Fig.~\Ref{fig:Introduction}c), approaching a level of saturation at 25\% where homology becomes challenging to detect \cite{Pearson2013}. The conserved amino acid residues are mostly associated with the canonical binding groove (Extended Data Fig.~\Black{1}a), consistent with the prevailing scientific perspective \cite{McGriff2023}, or are localized at the hydrophobic core of the protein.
As solely the human and murine homologs bear experimentally acquired structures (e.g. PDB: 6QFM, 2ROC), we used two structure prediction models, AlphaFold \cite{Jumper2021} and RosettaFold \cite{Baek2021}, to compute the structures for the remaining homologs (Extended Data Fig.~\Black{1}b). In comparison to their experimental equivalents, we found conserved topologies (TM scores $\geq$ 80\%\cite{Zhang2005,Xu2010}, Fig.~\Ref{fig:Introduction}d) and only small spatial differences between the predicted protein backbone  (RMSD $\leq$ 2.5~\AA, Fig.~\Ref{fig:Introduction}e). A subtle correlation between inferior structural conservation and increased divergence time became visible. Nevertheless, the predictions show that, although sequences might differ strikingly, MCL-1 structure did not substantially change over a long evolutionary time scale \cite{Chothia1986}.

The primary function of MCL-1, i.e., the ability to strongly bind the BH3 domain in its binding pocket, which makes it a pivotal anti-apoptotic regulator, is also conserved. We experimentally determined MCL-1's binding affinity for a uniform PUMA BH3 ligand (bearing mutations for crosslinking, see extended Data Fig.~\Black{2}), with $K_D$ values ranging from 100~nM to 1~$\mu$M for most homologs. We detected a weak correlation of $\log K_D$, which refers to the binding free energy, with evolutionary divergence time (Fig.~\Ref{fig:Introduction}f). Notably, homologs from both \textit{Hydra vulgaris} and \textit{Homo sapiens}, separated by an evolutionary distance of over 700~million years, bound the same ligand with comparable affinities ($K_{D,Hydra}$~=~220~nM, $K_{D,Homo}$~=~480~nM). Given its critical function as a 'life/death switch'\cite{Adams2007} in numerous animal species, this result confirms that MCL-1 indeed exhibits a high degree of structural and functional conservation, manifesting in minor differences at the molecular level. 

MCL-1's role as a prime example of structural and functional conservation raises the question of whether the dynamics of the protein are also conserved. Are the nanosecond processes occurring in human MCL\hbox{-}1 also present in \textit{Hydra vulgaris} MCL\hbox{-}1?

\section*{Conservation of protein dynamics}
\begin{figure*}[t]
	\centering
	\includegraphics[clip, trim=0cm 0cm 0cm 0cm, width=1\textwidth]{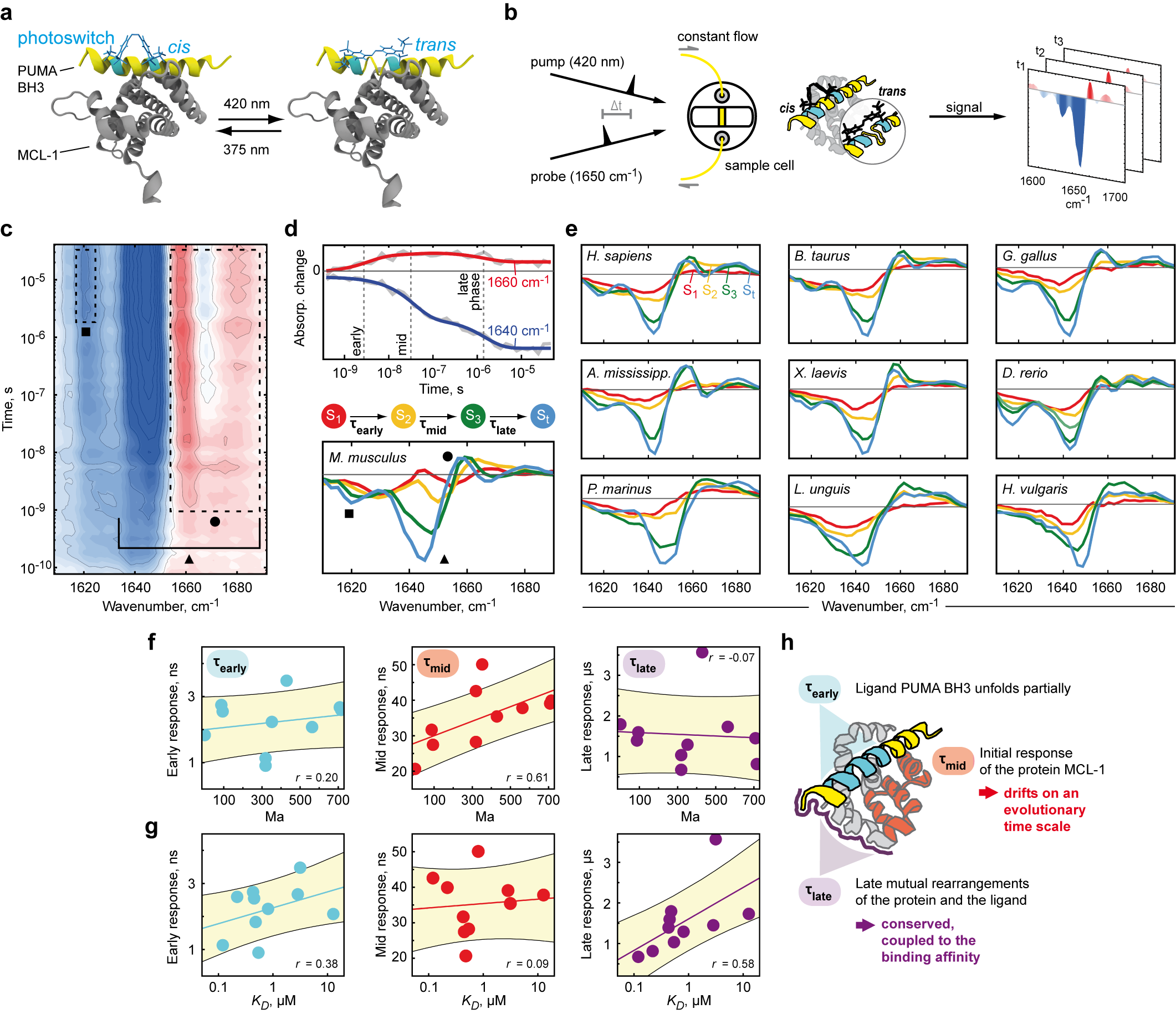}
	\caption {The conservation of ultrafast protein dynamics in MCL-1. (a) The protein MCL-1 in complex with the photoswitchable PUMA BH3 peptide. (b) Transient infrared spectroscopy of the photo-perturbed MCL-1/PUMA BH3 complex results in kinetic footprints for all homologs, exemplarily displayed in (c) for \textit{Mus musculus}. The symbols serve as reference points for explanations in the main text. (d) Three dominating phases of increased dynamic activity are assessed (early-, mid-, late phase; dashed lines). Global multiexponential fitting with three time constants yields fits (red/blue) that cover the raw data (grey) well. Evolution-associated difference spectra\cite{VanStokkum2004,Buhrke2020} (lower panel) were calculated for state S$_{1}$ (red), S$_{2}$ (yellow), S$_{3}$ (green), and for S$_{t}$ (blue) with time constants $\tau_{early}$, $\tau_{mid}$, and $\tau_{late}$. (e) The difference spectra of all homologs display a high degree of similarity. (f) Time constants of increased dynamic activity $\tau_{early}$, $\tau_{mid}$, and $\tau_{late}$ against evolutionary divergence in million years, Ma. (g) Time constants $\tau_{early}$, $\tau_{mid}$, and $\tau_{late}$ against MCL-1's affinity for PUMA BH3. Data in (f) and (g) are displayed with linear fits $\pm$ standard deviation (yellow) and correlation coefficients $r$ (Pearson). (h) Isotope labeling (Extended Data Fig.~\Black{5}) helped to separate the signal contribution of MCL-1 and PUMA BH3 spatially and temporally. The time constants were assigned to dynamic processes in the protein complex (schematic overview).}\label{fig:ProteinDynamics}
\end{figure*}

To examine the impact of extremely slow evolutionary processes on the fast-paced protein dynamics of MCL-1, we used transient infrared spectroscopy in combination with a photoswitchable azobenzene moiety that is covalently bound to the PUMA BH3 ligand (Fig.~\ref{fig:ProteinDynamics}a, Methods). In its \textit{cis}-state, the crosslinked photoswitch additionally stabilizes the ligand inside the binding pocket (Extended Data Fig.~\Black{2}m). Conversely, the light-induced transition from the \textit{cis}- to the \textit{trans} configuration leads to a reduction in $\alpha$-helicity (Extended Data Fig.~\Black{2}n),  indicating a destabilization of PUMA BH3.

Considering a time frame from pico- to microseconds\cite{Helbing2023}, we studied the protein dynamics in a pump-probe experiment where the \textit{cis}-to-\textit{trans} isomerization of the photoswitch is triggered by an ultrashort UV/VIS laser pulse at 420~nm and the protein vibrational spectrum is probed in the mid infrared region around 1650~cm$^{-1}$ (Fig.~\ref{fig:ProteinDynamics}b). In this spectral region, C=O stretch vibrations of the protein backbone can be observed. Negative (blue) and positive (red) absorption changes serve as indicator for structural alterations\cite{barth02}. 
We obtained homolog-specific kinetic footprints for the ten investigated species (Extended Data Fig.~\Black{3}, exemplified for \textit{M. musculus} in Fig.~\ref{fig:ProteinDynamics}c). Analogous to fossil footprints -- the paleontologic counterpart --, the kinetic footprints display comparable elements. All of them are similarly shaped, displaying a blue shift of a band at 1645~cm$^{-1}$, which reveals a negative bleach towards a new (positive) band at 1675~cm$^{-1}$ (Fig.~\ref{fig:ProteinDynamics}c, triangle). The signal appears within the low nanosecond time frame for all of the homologs and can be attributed to $\alpha$-helix unfolding\cite{Barth2007,Huang2002,Heckmeier2022}. More strikingly, the kinetic footprints exhibit diverging, species-specific details, which are particularly well visible for the spectral feature between 1655~cm$^{-1}$ and 1685~cm$^{-1}$ (Fig.~\ref{fig:ProteinDynamics}c, circle), and a late negative feature at 1620~cm$^{-1}$ forming at around 100~ns (Fig.~\ref{fig:ProteinDynamics}c, square). The first-mentioned feature (circle) is especially pronounced for mammalian/avian/reptilian homologs (Extended Data Fig.~\Black{3}a-e), but loses its distinct appearance more and more for species with higher evolutionary divergence (\textit{P.~marinus}, \textit{L.~unguis}, \textit{H.~vulgaris} (Extended Data Fig.~\Black{3}h-j), displaying a solitary, less emphasized maximum at 1660~cm$^{-1}$. Furthermore, the kinetic footprints of the non-\textit{Gnathostomata} \textit{P.~marinus}, \textit{L.~unguis}, and \textit{H.~vulgaris}, lack the late negative feature at 1620~cm$^{-1}$.  

All kinetic footprints are dominated by three phases of dynamic activity, an early-, mid-, and late phase, where the intensity of spectral features grows or decreases significantly (exemplified for \textit{M. musculus} in Fig.~\ref{fig:ProteinDynamics}d, for all other species in Extended Data Fig.~\Black{4}). To fathom these three dynamic processes and their corresponding time constants, we analyzed the kinetic footprints with global multiexponential fitting (Fig.~\ref{fig:ProteinDynamics}d, details in Methods section)\cite{VanStokkum2004,Lorenz2006,Buhrke2020,Heckmeier2022}. Our analysis demonstrates that there are four states of molecular rearrangement upon photo-perturbation, populated with time constants $\tau_{early}$, $\tau_{mid}$, and $\tau_{late}$. 
The time intervals in which the three observed processes take place are very narrow for the ten homologs we investigated, evidencing that not only the structure and function of MCL-1 are conserved across a wide and diverse range of today's living animals (Fig.~\ref{fig:Introduction}) but also the underlying protein dynamics (Fig.~\ref{fig:ProteinDynamics}e). This stands in stark contrast to the significant alterations that we observe for the primary structure of the protein homologs (Fig.~\ref{fig:Introduction}c). When we plot the time constants against an evolutionary time scale (Fig.~\ref{fig:ProteinDynamics}f), we find that the processes populated with $\tau_{mid}$, correlate with the evolutionary divergence. In contrast, we did not detect similar protein dynamic drifts for the other two time constants, showing an absence of correlation of early- and late protein response with evolutionary divergence. On the other hand, if the time constants are plotted in dependence of the experimental binding affinities, it becomes evident that the processes populated with $\tau_{late}$ are strongly correlated with the protein's affinity (Fig.~\ref{fig:ProteinDynamics}g). 

We specified which parts of the protein complex contribute to which process by recording kinetic footprints for $^{13}$C-$^{15}$N-labelled MCL-1 of \textit{M.~musculus} in complex with non-labelled PUMA BH3 (Extended Data Fig.~\Black{5}). Separating in this way the protein from the peptide response, we found that  time constant $\tau_{early}$ (=~0.9-3.5~ns) can be attributed to the $\alpha$-helical unfolding of the PUMA BH3 peptide.(Fig.~\ref{fig:ProteinDynamics}h)
The time constant $\tau_{mid}$ (=~21-50~ns) corresponds to spectral features which shift $\approx$~50~cm$^{-1}$ for isotope labelled MCL-1 (Extended Data Fig.~\Black{5}), and can thus be traced back to a initial response of MCL-1, that potentially allows to rearrange and cope with the conformational destabilization originating from the binding pocket. Apparently, mammalian homologs exhibit an earlier MCL-1 adaption upon destabilization than non-mammalian/vertebrate species and again other non-\textit{vertebrata} (Fig.~\ref{fig:ProteinDynamics}f, red). Finally, the terminal time constant $\tau_{late}$ (=~0.7-3.6~$\mu$s) corresponds to mutual rearrangements in the whole complex. The results are in line with previous observations for the isotope labelled human MCL-1/BIM complex \cite{Heckmeier2022}.

From our results, one might speculate whether \mbox{MCL-1's} initial response ($\tau_{mid}$) has met with less selective pressure in the past, causing it to drift. This hypothesis is supported by the absence of any discernible correlation between $\tau_{mid}$ and the protein affinity (Fig.~\ref{fig:ProteinDynamics}g), implying that the function of the protein is seemingly not entangled with this dynamic process. In contrast, a robust correlation between $\tau_{late}$ and the $K_{D}$(Fig.~\ref{fig:ProteinDynamics}g) indicates that the late mutual rearrangements of MCL-1 and PUMA BH3 (in the microsecond regime) are connected to the function of the protein. From our observations, it seems that the relationship between the late dynamic response and the protein affinity is conserved and cannot be inferred from the evolutionary separation of species; other factors must be at play. 
Irrespective of how to evaluate the given correlations, what remains truly remarkable is that our results provide an unprecedented opportunity to gain insights into the speed and extent of the impact of evolution on dynamical processes.

\section*{Conclusion}
MCL-1 is a critical player in apoptosis\cite{Youle2008}, not only in human beings but also in a great variety of animals\cite{Aouacheria2015}. By experimentally studying ten MCL-1 homologs and their interactions with a photo-switchable ligand PUMA BH3, we gained valuable insights into the dynamics of the proteins on a broader evolutionary time scale. Using time-resolved infrared spectroscopy, we successfully recorded the kinetic footprints of the MCL-1/PUMA BH3 complex and analytically compared them -- similar to bones, skulls, and footprints in the classic field of paleontology\cite{Stern1983,Olsen1998}, or protein structures and genetic information in its molecular form\cite{King1975,Reich2010,VanderValk2021}.

Our findings reveal a remarkable degree of conservation for the protein dynamics across the homologs, highlighting the importance of these processes in preserving their anti-apoptotic function over a span of nearly a billion years. Of particular interest is the correlation we observed between one of these ultrafast processes and the evolutionary divergence among the protein homologs, a drift in protein dynamics in the nanosecond range. This discovery challenges the prevailing focus on resolving protein structures\cite{Hadzipasic2020} and dynamics in equilibrium\cite{Jemth2018} or analyzing genomic data\cite{VanderValk2021} to understand evolution. Instead, our work highlights the importance of considering nanosecond protein dynamics as a crucial factor in unraveling the evolutionary history of these proteins. With this approach, we build a bridge between the shortest (1~ns~=~10$^{-9}$~s) and the largest conceivable timescales in living matter (300~Ma~$\approx$~10$^{16}$~s).

Overall, our study defines a starting point for exploring the dynamics of countless other proteins with varying degrees of conservation. By investigating different systems that are more or less conserved, we can gain valuable insights into the extent of evolution's impact on nanosecond processes, and how these rapid processes translate to slow-paced protein function. 

\makeatletter
\def\@biblabel#1{(#1)}
\makeatother

\section*{Methods}
    \subsection*{Phylogeny and Bioinformatics}
    From the countless species in the tree of life we chose
    ten MCL-1 homolog sequences (Fig.~1b). Alongside the \textit{Homo sapiens} homolog, our selection encompasses a variety of species, including mammalian (\textit{Mus musculus}, \textit{Bos taurus}), avian (\textit{Gallus gallus}), reptile (\textit{Alligator mississippiensis}), amphibian (\textit{Xenopus laevis}), bony fish (\textit{Danio rerio}), and other farther related eumetazoan homologs (\textit{Petromyzon marinus}, \textit{Lingula unguis}). Notably, we also incorporated a homolog from \textit{Hydra vulgaris}, one of the most distantly related organisms known to exhibit BCL-2 regulated apoptosis\cite{Lasi2010}. The curated selection represents species whose last common ancestors existed at quasi-equidistant intervals spanning nearly a billion years of evolutionary history. 
    
    To assemble our dataset, we accessed amino acid sequences from the Uniprot database. All entries shared the common identifier ``MCL-1" in their title or description. The amino acid sequence for \textit{P. marinus} was added to the selection with the help of Jeramiah Smith (gene on Chr52: 9161036..9167581, + strand; annotated: PMZ\_0059412-RA) \cite{Timoshevskaya2023}. The sequences were aligned to the human variant (soluble domain, $\Delta$N-$\Delta$C aa 171–327\cite{Czabotar2007}) and harmonized in length ($\approx$150-160 aa) (Extended Data Fig.~\ref{fig:EDBioinfo}a). The chosen sequences (refer to Extended Data Tab.~\ref{tab:Seq_Divtimes}) were initially controlled for by predicting their structure using AlphaFold and RosettaFold (see below) and aligning them with experimental structures from \textit{Homo sapiens} and \textit{Mus musculus}. From the sequences, we generated a multiple sequence allignement using Clustal Omega (EMBL-EBI)\cite{Sievers2011} (Fig.~\ref{fig:EDBioinfo}a).
    
    The phylogeny in Fig.~1b was obtained from the \textit{TimeTree} database (http://timetree.org)\cite{Kumar2022}. It was not computed from the investigated MCL-1 sequences. In contrast, the given phylogeny was constructed from median and adjusted divergence times which were estimated by \textit{TimeTree} based on values from an abundance of published studies. The divergence times, always related to \textit{H. sapiens} and tabulated in Extended Data Tab.~\ref{tab:Seq_Divtimes} alongside their corresponding confidence interval, reflect the  most current scientific understanding (June 2023). In figures Fig.~1c-f, and Fig.~2f, the divergence time is given in million years, Ma.  

    The experimental structures of MCL-1/PUMA were retrieved from PDB (\textit{Mus musculus}: 2ROC; \textit{Homo sapiens}: 6QFM). In addition, we predicted the structures of all MLC-1 homologs with AlphaFold \cite{Jumper2021} and RosettaFold \cite{Baek2021}. We used ColabFold \cite{Mirdita2022} to generate AlphaFold-predicted structures, and the Robetta server \cite{Baek2021} for RosettaFold-predicted structures, both with default parameters. To estimate the structural similarity between all protein pairs, we performed an all-against-all alignment of the predicted structures and computed the TM score and root-mean-square deviation (RMSD) of each protein pair using TM-align \cite{Zhang2005}. For both AlphaFold and RosettaFold, we selected the top ranked structure out of the five predictions for downstream analyses. We evaluated the quality of the predicted structure using AlphaFold predicted local distance difference test (pLDDT), a per-residue confidence metric which estimates how well the predicted structure would match with an experimental one and which has been shown to be well-calibrated \cite{Jumper2021}. All our predicted structures have high average pLDDT values, ranging from 0.83 to 0.93, indicating good quality predictions.
    
    \subsection*{Protein preparation}
    Examining protein function and dynamics, we expressed ten different MCL-1 homologs using a \textit{Escherichia coli} BL21 expression strain (Fig.~\ref{fig:EDCD}a). Initially, the bacterial cells were transformed with a pET-30a(+) plasmid containing the corresponding MCL-1 homolog gene, using electroporation. Positive clones were selected through Kanamycin resistance.
    For standard expression, bacterial cultures were cultivated in lysogeny broth medium until reaching an optical density of OD$_{600}$~=~0.6. The expression was induced by adding 700~$\mu$M IPTG, followed by incubation at 30~°C for 20 hours. Cell harvest was carried out through centrifugation (3000 x\textit{g}). 
    In order to generate heavy, uniformly $^{13}$C$^{15}$N-labeled MCL-1, bacterial cultures were grown in minimal medium supplemented with solely heavy carbon and nitrogen sources. The cells were cultivated to an OD$_{600}$ of 0.6, induced with 1~mM IPTG, and then further incubated at 30~°C. The expression was stopped after 4 hours with cell harvest as described above. 
    Cell lysis was achieved by subjecting the harvested cells to sonication (20~kHz, 4 x 1~min pulses). The lysed cell suspension was purified using Ni-affinity chromatography and a His$_6$-Tag located at the N-terminus of the protein. Purification was carried out under native conditions. The N-terminal His$_6$-Tag was removed by 3C protease cleavage. 
    Throughout this study, all analytical procedures were performed in a sample buffer composed of 50 mM Tris (pH 8) and 125 mM NaCl. Mass spectrometry was used to assess the protein's integrity and sample purity. For long-term storage, the samples were kept at -80~°C. 
    In total, we could express the homologs of ten species given in the main text (Extended Data~Fig.~\ref{fig:EDYield}). Under identical conditions, however, we could not express \textit{Ornithorhynchus anatinus}, \textit{Orchesella cincta}, and \textit{Acanthaster planci} homologs at adequate concentrations.
    
    \subsection*{Peptide preparation}
    PUMA BH3 (EEQWAREIGAQLRCMADDLNCQYERV) was synthesized using solid-state peptide synthesis on a Liberty 1 peptide synthesizer (CEM corporation, Matthews, NC, USA). In this study, the peptide was deliberately modified by introducing two mutations -- replacing Arg143 and Ala150 with Cys residues -- compared to the native mammalian version. These Cys residues were incorporated distal to the hydrophobic binding interface, to enable the covalent linkage of a photoswitchable azobenzene moiety.
    To achieve this linkage, the watersoluble photoswitch (3,3'-bis(sulfonato)-4,4'-bis(chloroacetamido)azobenzene)\cite{Zhang2003} and the peptide with reduced Cys residues were together incubated in a 20 mM Tris (pH 8.5) at a temperature of 50~°C, under continuous stirring for a duration of 20 hours.
    Hereafter, the reaction product underwent purification using both anion exchange and reversed-phase chromatography (C18 10$\mu$m) to isolate the successfully linked peptide. For final preparation, the buffer of the isolated linked peptide was exchanged through dialysis against the sample buffer (50~mM Tris pH~8, 125~mM NaCl). The linkage's success, as well as the peptide's purity and integrity, were controlled via mass spectrometry.

    \subsection*{Circular dichroism spectroscopy}
    The expressed MCL-1 homologs have in common that they contain eight $\alpha$-helical elements \cite{Czabotar2007}, and exhibit a circular dichroism spectrum that is typical for $\alpha$-helical structures (Extended Data Fig.~\Black{2}b, yellow).
    In contrast, their peptide ligand PUMA BH3 is intrinsically disordered in isolation\cite{Hinds2007} (Extended Data Fig.~\Black{2}b, grey). When in complex with MCL-1, PUMA BH3 assumes an $\alpha$-helical shape (Extended Data Fig.~\Black{2}b, black). 

    We utilize circular dichroism spectroscopy to accomplish two distinct objectives: (i) to evaluate the $\alpha$-helical content of the MCL-1 and PUMA BH3 complex at a constant concentration, thereby assessing whether they are correctly folded, and (ii) to generate binding curves and determine dissociation constants ($K_D$) for all analyzed MCL-1 homologs. To record binding curves and assess the $K_D$ values, we exploited the nature of PUMA BH3 which is intrinsically disordered in solution and only exhibits an $\alpha$-helical secondary structure when bound by MCL-1's binding groove. Hence, for an increasing concentration of bound PUMA BH3, and a constant concentration of MCL-1, the $\alpha$-helical content added by titration reflects the fraction of bound peptide.  
    
    For the first aspect (i), a quartz glass cuvette with a 1~mm path length was employed, and the sample concentration was maintained at 20~$\mu$M. We measured the spectrum between 200-260~nm at room temperature. Hereby, we examined the $\alpha$-helical content of the MCL-1 and PUMA BH3 complex which served as a control to for their correct structural conformation (displayed in Extended Data Fig.~\ref{fig:EDCD}b).
    
    For the second aspect (ii), MCL-1 was brought to a concentration of 2~$\mu$M. A quartz glass cuvette with a path length of 1~cm was used, and continuous stirring was maintained during the spectroscopic measurements at room temperature. To record the binding curves, we titrated both the linked and unlinked forms of the PUMA BH3 peptide to the MCL-1 homolog, offering a comprehensive understanding of the binding affinity of photoswitchable and non-photoswitchable complexes. The circular dichroism was recorded at 222~nm as a function of increasing PUMA BH3 concentration.
    In both scenarios (i) and (ii), measurements involving the photoswitchable PUMA BH3 were conducted for both the \textit{cis}-state (achieved through illumination with a 375~nm laser) and the dark-adapted \textit{trans}-state.  
    
    By recording the $\alpha$-helical content at 222~nm as a function of increasing PUMA BH3 concentration, we received binding curves for all MCL-1 homologs. In order to calculate the dissociation constant $K_D$, we fitted the data to a two-state binding equilibrium\cite{Jankovic2019,Jarmoskaite2020}:
    
    \begin{equation}\label{BindingAffinity}
    K_D = \frac{([M]-[MP])\times([P]-[MP])}{[MP]}
    \end{equation}
    
    where [M] is the initial concentration of the MCL-1, [P] is the initial concentration of PUMA BH3 given to the solution, and [MP] is the concentration of the protein-peptide complex. For a constant [M]~=~2~$\mu$M and a variable [P], the fraction of bound peptide can be understood as:

    \begin{equation}\label{FractionBound}
    \resizebox{0.4\textwidth}{!}{$\rm{Fraction~bound} = \frac{([M]+[P]+K_D)-\sqrt{([M]+[P]+K_D)^2-4\times[M]\times[P]}}{2\times[M]}$}
    \end{equation}
    
    The covalently bound photoswitch in the \textit{cis}-state stabilizes PUMA BH3 inside the binding pocket (Fig.~1a), with significantly lower $K_D$ values for all of the homologs (Extended Data Fig.~\ref{fig:EDCD}m). For PUMA BH3 in the \textit{cis} state, \textit{Homo sapiens}, \textit{Bos taurus}, and \textit{Alligator mississippiensis} homologs showed the highest affinities, with  $K_D$ values in the low nanomolar regime ($<$10~nM), a region that was classified as physiological for wild type PUMA \cite{Kale2018}. Switching the photoswitch from its \textit{cis}- to its \textit{trans} configuration results in a loss in $\alpha$-helicity (Extended Data Fig.~\ref{fig:EDCD}n) and in the destabilization of PUMA BH3. 
    
    \subsection*{Transient infrared spectroscopy}
    MCL-1 and PUMA BH3 were mixed in a 1:1 ratio prior to the spectroscopic experiment. To ensure high signal strength in transient infrared spectroscopy, both the protein and peptide were brought to concentrations of 600~$\mu$M each in the final sample. The overall sample volume was 800~$\mu$L.
    Considering concentrations $>$500~$\mu$M, it is expected that the PUMA BH3 peptide will be predominantly bound within MCL-1's binding pocket, as illustrated in Extended Data Fig.~\ref{fig:EDCD}. To exclude H$_2$O from spectroscopic experiments, the employed buffer was exchanged to a corresponding buffer containing D$_2$O. Stringent precautions were taken to avert H$_2$O contamination by preserving the sample within a nitrogen atmosphere devoid of water vapor. 
    
    For pump-probe measurements, a pair of electronically synchronized 2.5~kHz Ti:sapphire oscillator/regenerative amplifier femtosecond laser systems (Spectra Physics) were employed,  offering a maximal delay of 45~$\mu$s \cite{Bredenbeck2004, Helbing2023}. One of these laser systems, featuring frequency-doubled pulses (420~nm, 3~$\mu$J per pulse, focused to an approximate beam diameter of 140~$\mu$m within the sample, and stretched to $\sim$60~ps to minimize sample deposition on the sample cell windows), were used to induce the \textit{cis} to \textit{trans}-isomerization of the photoswitch. The second laser system was applied to generate infrared probe pulses via an optical parametric amplifier (100~fs, spot size 110~$\mu$m, center wavenumber 1660~cm$^{-1}$). 
    To ensure a consistent sample environment, the sample was continuously circulated within a closed-cycle flow cell system. This system consisted of a reservoir and a CaF$_{2}$ cell featuring a 50~$\mu$m optical path length. Before entering the measurement cell, the sample was irradiated in a pre-illumination step using a 375~nm continuous wave diode laser (90~mW, CrystaLaser), in order to optimally prepare the sample with $>$85\% in the \textit{cis}-state.

    \subsection*{Data analysis}
    From time resolved infrared measurements, we obtained kinetic footprints in the form of 2D data sets $d(\omega_i,t_j)$ as a function of probe frequency $\omega_i$ and pump-probe delay time $t_j$ (Fig.~2c,  Extended Data Fig.~\ref{fig:EDFootprints}). For each homolog, we subjected the 2D dataset to a global multiexponential fitting \cite{Buhrke2020}, operating under the premise that the investigated system can be understand as interconverting discrete states with time-invariant spectra.
    
    We employed multiexponential functions with amplitudes $a(\omega_i,\tau_k)$ and a global set of time constants $\tau_{k}$ for fitting the experimental data\cite{Hobson1998,Kumar2001,Lorenz2006}:
    \begin{equation}
    d(\omega_i,t_j)=a_0(\omega_i)-\sum_{k} a(\omega_i,\tau_k)e^{-t_j/{\tau_k}}.
    \label{LDA}
    \end{equation}   
    
   The time constants $\tau_{k}$ were treated as free fitting parameters, with the constraint of a minimal number of exponential terms. 
   Based on observations with similar systems\cite{Heckmeier2022,Heckmeier2023} we excluded data before 300 ps, to prevent the influence of the pump pulse (60~ps pulse length), which results in a strong ``heat signal'' at 100~ps, induced by azobenzene photoisomerization, which can be universily observed for this kind of experiment \cite{Heckmeier2022,Heckmeier2023,Bozovic2020b}. Three time constants $\tau_{early}$, $\tau_{mid}$, and $\tau_{late}$ were needed to adequately fit the data, dissecting the dynamic response in an early-, mid-, and late phase (Extended Data Tab.~\ref{tab:Tau}).  The one exception is \textit{D.~rerio}, which required a fourth time constant  $\tau_{D.rerio}$ = 300~ns to adequately fit the data.
    
    Under the assumption of a sequential, unidirectional process with four states S$_1$, S$_2$, S$_3$, S$_t$ and three time constants $\tau_{early}$, $\tau_{mid}$, $\tau_{late}$ connecting them, we calculated concentration profiles for each state as well the corresponding evolution-associated difference spectra\cite{VanStokkum2004}, which are depicted in Fig.~2d,e. Commencing with state S$_{1}$, all subsequent evolution-associated difference spectra exhibited a blue shift from 1645~cm$^{-1}$ to around 1675~cm$^{-1}$. Equally to our observations for the raw data, a very distinct positive feature at 1660~cm$^{-1}$ was detected in the evolution-associated difference spectra of the latest two states (Fig.~2d,e, green and blue).
    
    With the help of isotope labeling\cite{Haris1992} (Fig.~\ref{fig:EDIsotope}), we could assign this distinct spectral maximum, populated with $\tau_{mid}$, to the initial response of MCL-1 upon photo-destabilization of its ligand. The early response at $\tau_{early}$ exclusively originates from the unfolding of PUMA BH3. The terminal, late response at $\tau_{late}$ results from mutual, heterogeneous rearrangements in MCL-1 and PUMA BH3. 

\section*{Abbreviations}
BH3, BCL-2 Homology Domain~3; MCL-1, Myeloid Cell Leukemia~1.

\section*{End notes}

\noindent\textbf{Acknowledgements}
We thank Markus B. Glutz for the synthesis of the peptide and Serge Chesnov from Functional Genomics Center Zurich for their work on the mass spectrometry and amino-acid analysis. We thank Jeramiah Smith, University of Kentucky, who provided us with the sequence of the \textit{Petromyzon marinus}. The work has been supported by the Swiss National Science Foundation (SNF) through the Sinergia grant CRSII5\_213507.\\

\noindent\textbf{Author contributions} P.J.H conceived the study. P.J.H. selected homologs and gathered sequence information, performed microbiological work, protein expression and purification, azobenzene crosslinking, and the purification of the crosslinked peptide. P.J.H. performed binding studies with circular dichroism spectroscopy. J.R. and P.J.H performed time-resolved infrared spectroscopic measurements. C.R. performed structure predictions and computed structural similarity for the homologs. P.H. conceived and built the experimental pump probe setup. P.H. developed the analysis tools for the experimental data evaluation. P.J.H. analysed the experimental data, with input from P.H. and J.R.. P.H. supervised the project and acquired the funding. P.J.H. prepared the figures. P.J.H. wrote the manuscript with strong contribution from P.H. and minor contribution from all other authors.\\

\noindent\textbf{Competing interests} \\
The authors declare no competing financial interest.\\

\noindent\textbf{Additional Information} \\

\noindent\textbf{Correspondence and requests for materials}
should be addressed to Philipp J. Heckmeier.



\noindent\textbf{Data Availability:} The data that support the findings of this study are openly available in Zenodo (the link will be provided at the proof stage.)\\

\bibliography{Library}

\setcounter{figure}{0}  
\renewcommand{\figurename}{Extended Data Fig.}
\renewcommand{\tablename}{Extended Data Table}

     \begin{figure*} [p]
    	\centering
    	\includegraphics[clip, trim=0cm 0cm 0cm 0cm, width=1\textwidth]{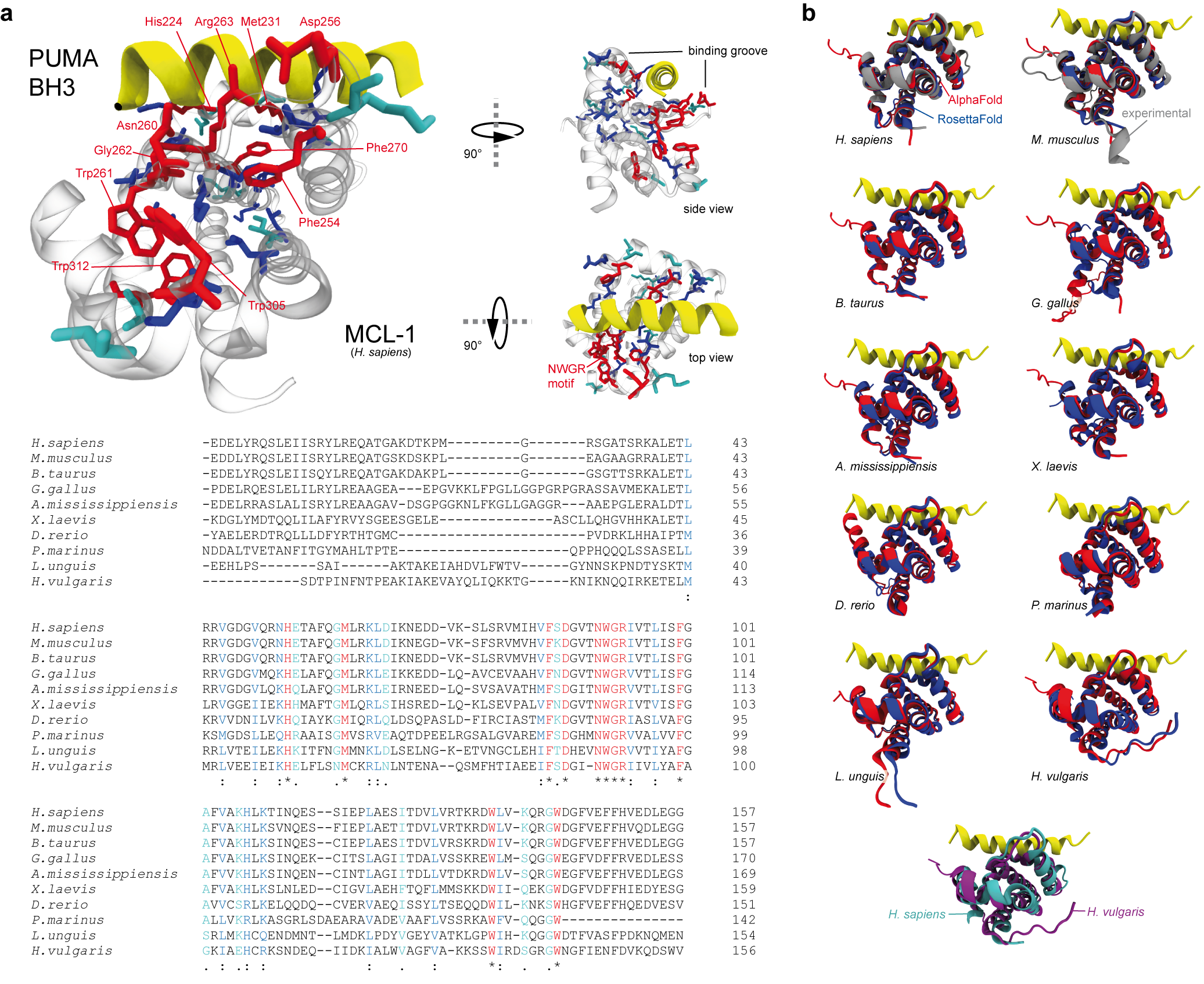}
    	\caption{(a) Location of the conserved residues in MCL-1 (PDB: 6QFM) according to a multiple sequence alignment of MCL-1 homolog selection. The residues in the structure are given with their position in \textit{Homo sapiens}. The alignment was generated using Clustal Omega (EMBL-EBI)\cite{Sievers2011}. "*", red: positions with a completely conserved residue. ":", blue: positions with high conservation between groups of strongly similar properties. ".", cyan: conservation between groups of weakly similar properties. The highlighted NWGR motif is characteristic for the BH1 domain of MCL-1\cite{McGriff2023} and present in all investigated homologs. (b) Structural alignment of predicted AlphaFold (red) and RosettaFold (blue) structures for the investigated MCL-1 homologs. Experimental structures (grey) were only available for \textit{Homo sapiens} and \textit{Mus musculus}. The complexed PUMA BH3 is displayed in yellow and was taken from the corresponding experimental structures. For homologs without experimental structures, the structure of PUMA BH3 originally from \textit{Mus musculus} is displayed representatively. The structures at the bottom show a comparison between the experimental MCL-1 structure of \textit{Homo sapiens} (6qfm, cyan) and the AlphaFold predicted structure of \textit{Hydra vulgaris} (purple), two species which exhibit a evolutionary separation by more than 700 million years.}
            \label{fig:EDBioinfo}
    	\addcontentsline{toc}{subsection}{Figure~\ref{fig:EDBioinfo}}
            \end{figure*}
\newpage 
    
            \begin{figure*} [p]
    	\centering
    	\includegraphics[clip, trim=0cm 0cm 0cm 0cm, width=1\textwidth]{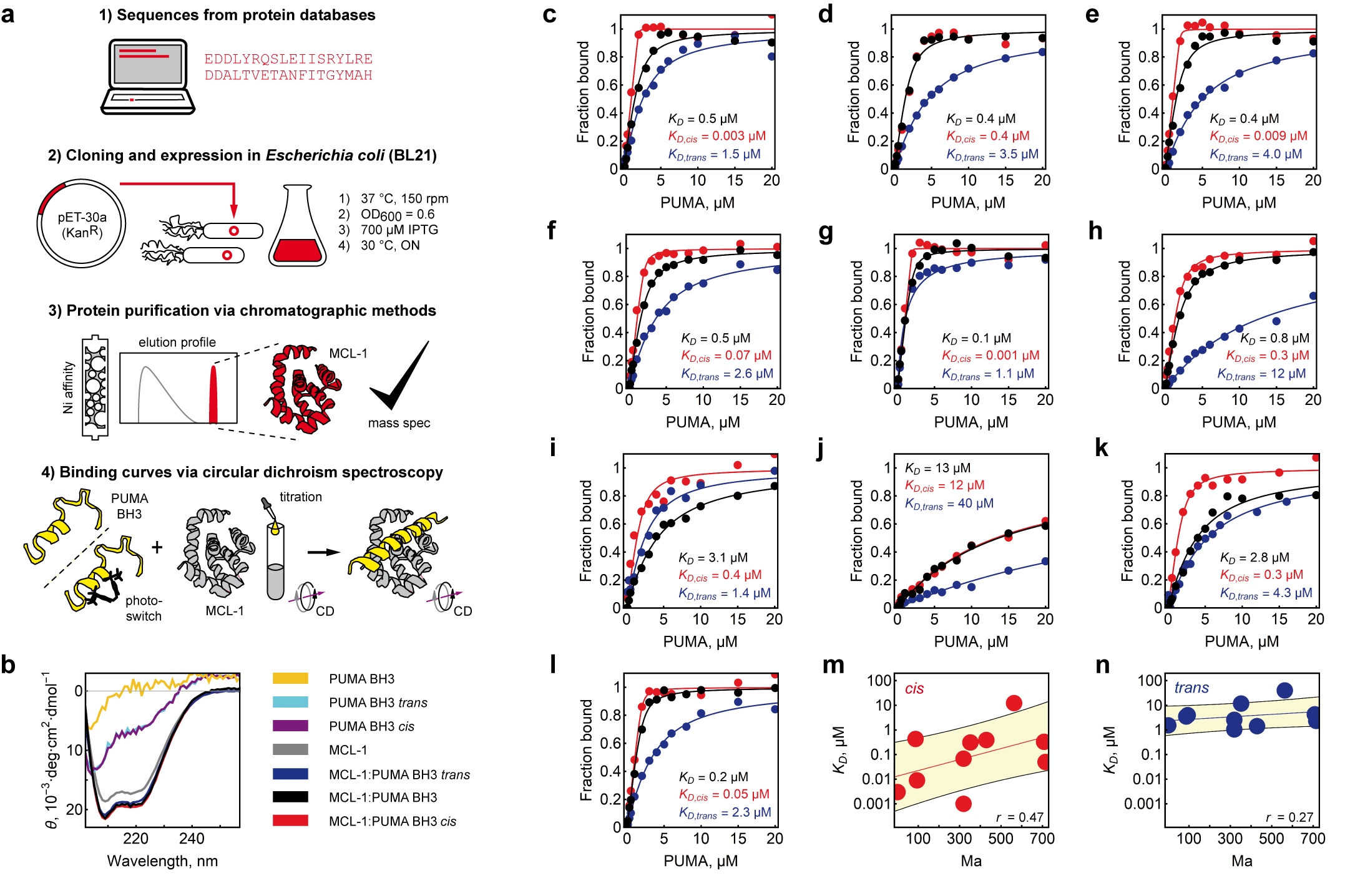}
    	\caption{The binding affinities of ten MCL-1 homologs. (a) Generation of ten MCL-1 homologs for spectroscopic experiments. (b) Circular dichroism spectroscopy of the MCL-1/PUMA BH3 complex. PUMA BH3 is intrinsically disordered and only assumes an $\alpha$-helical shape when bound to MCL-1. When linked to an azobenzene photoswitch, PUMA BH3 is destabilized in the \textit{trans}-state of the photoswitch moiety. Binding curves for (c)~\textit{Homo sapiens}, (d)~\textit{Mus musculus}, (e)~\textit{Bos taurus}, (f)~\textit{Gallus gallus}, (g)~\textit{Alligator mississippiensis}, (h)~\textit{Xenopus laevis}, (i)~\textit{Danio rerio}, (j)~\textit{Petromyzon marinus}, (k)~\textit{Lingula unguis}, and (l)~\textit{Hydra vulgaris}. The binding affinity in \textit{cis}-state (m) and \textit{trans}-state against the evolutionary divergence in million years, Ma. Yellow, linear fit $\pm$ standard deviation; plots are displayed with the Pearson correlation coefficient $r$.} \label{fig:EDCD}
    	\addcontentsline{toc}{subsection}{Figure~\ref{fig:EDCD}}
        \end{figure*} 
\newpage
            \begin{figure*} [p]
    	\centering
    	\includegraphics[clip, trim=0cm 0cm 0cm 0cm, width=1\textwidth]{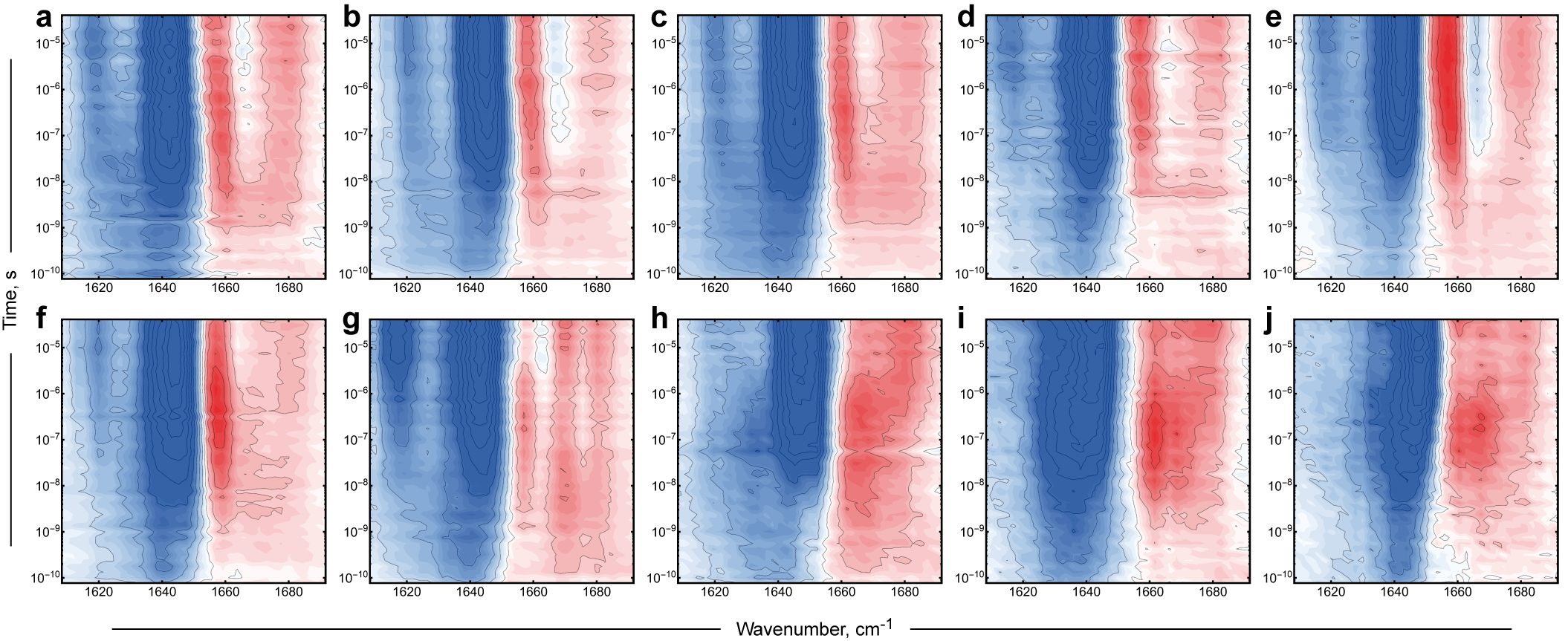}
    	\caption{Kinetic responses, serving as footprints, for (a) \textit{Homo sapiens}, (b) \textit{Mus musculus} (divergence time: 87~million years, Ma), (c) \textit{Bos taurus} (94~Ma), (d) \textit{Gallus gallus} (319~Ma), (e) \textit{Alligator mississippiensis} (319~Ma), (f) \textit{Xenopus laevis} (352~Ma), (g) \textit{Danio rerio} (429 Ma), (h) \textit{Petromyzon marinus} (563 Ma), (i) \textit{Lingula unguis} (708 Ma), and (j) \textit{Hydra vulgaris} (715 Ma). The kinetic footprints display comparable elements. We observe fine differences between closer related homologs and diverging, gradually emphasized features for farther related homologs, indicating an evolutionary timescale in which protein dynamic processes have changed.} \label{fig:EDFootprints}
    	\addcontentsline{toc}{subsection}{Figure~\ref{fig:EDFootprints}}
        \end{figure*} 
\newpage
            \begin{figure*} [p]
    	\centering
    	\includegraphics[clip, trim=0cm 0cm 0cm 0cm, width=0.6\textwidth]{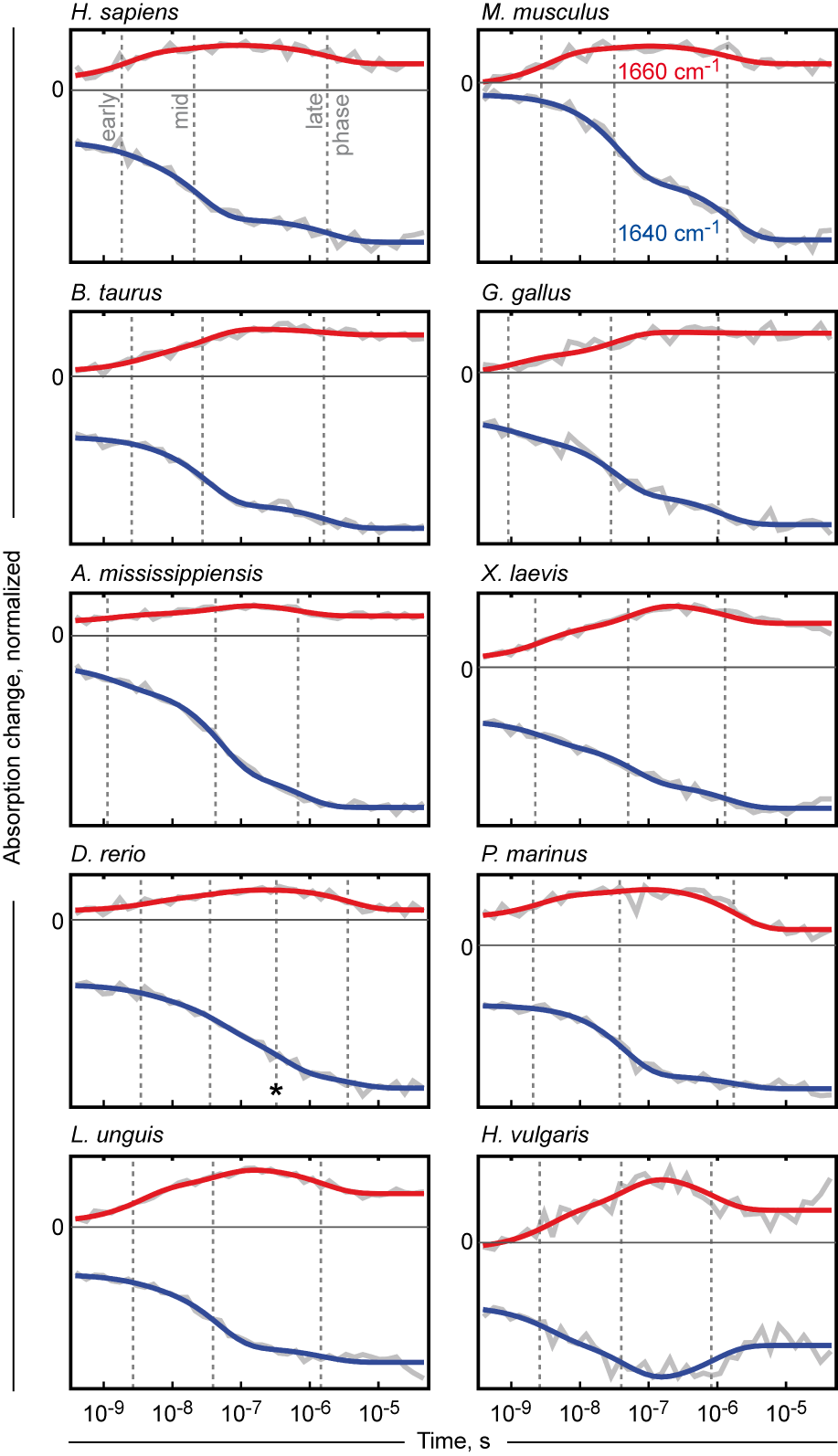}
    	\caption{The dynamic response upon photo-perturbation can be subdivided into an early-, mid-, and late phase, exemplified here for kinetic traces at 1640~cm$^{-1}$ and 1660~cm$^{-1}$. When analyzed with global multiexponential fitting and three time constants $\tau_{early}$, $\tau_{mid}$, and $\tau_{late}$ (given as dashed lines), the resulting fits (red/blue) are congruent with the time traces (grey). The one exception is \textit{D.~rerio}, which requires an additional fourth time constant  $\tau_{D.rerio}$ = 300~ns (*) to adequately fit the data.} \label{fig:EDFitting}
    	\addcontentsline{toc}{subsection}{Figure~\ref{fig:EDFitting}}
        \end{figure*} 
\newpage
            \begin{figure*} [p]
    	\centering
    	\includegraphics[clip, trim=0cm 0cm 0cm 0cm, width=0.4\textwidth]{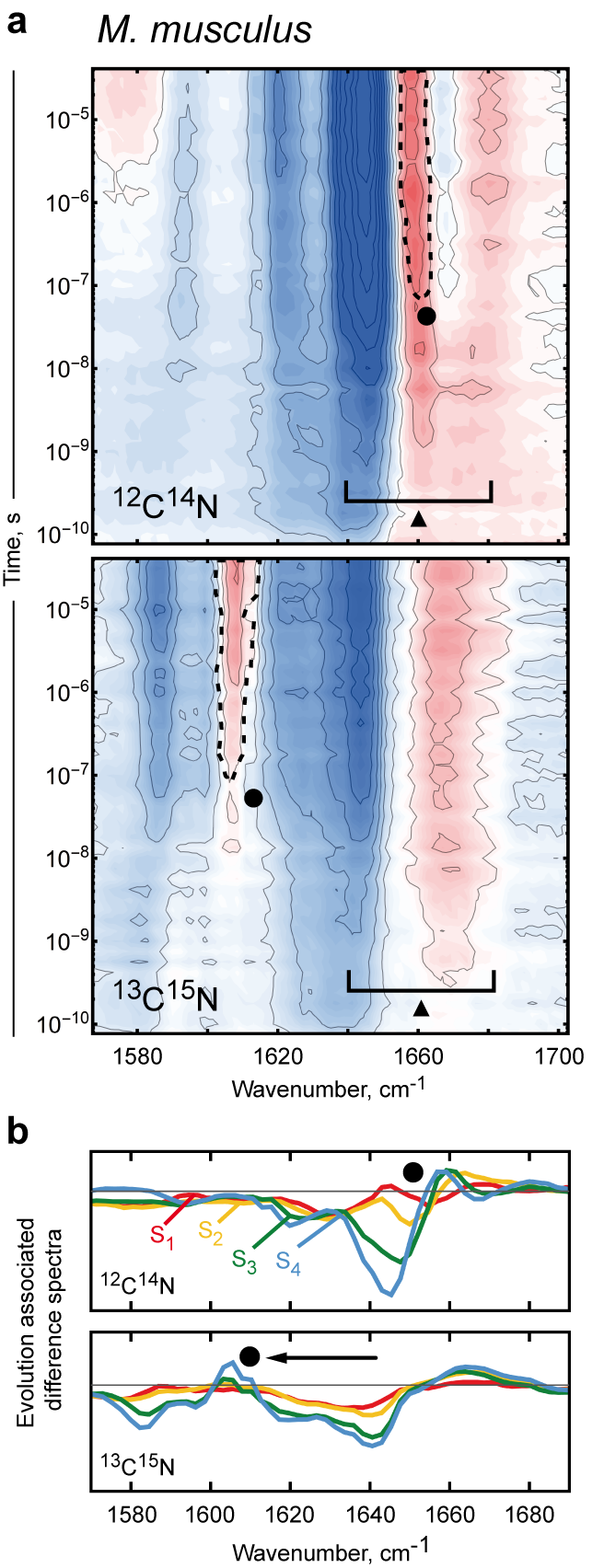}
    	\caption{Transient infrared spectroscopy with isotope labeled MCL-1/PUMA BH3 complexes. From ten homologs, \textit{M. musculus} was chosen as a representative, as $^{13}$C$^{15}$N-labeling is highly cost intensive. (a) Kinetic footprints of unlabeled samples (upper panel), and of samples with $^{13}$C$^{15}$N-labeled MCL-1 (lower panel). The early protein response (triangle) is not shifted for the labeled sample. The mid protein response (circle) manifests in a distinct sharp feature that shifts from 1660 cm$^{-1}$ to 1610 cm$^{-1}$ upon isotope labeling (dashed lines). Isotope labeling did not separate any spectral features at the late phase of the protein response. (b) Evolution-associated difference spectra (see Data analysis in the Methods section) display a relatively sharp positive band in state S$_{3}$ and S$_{4}$ (circle). In the non-labeled complex, the sharp maximum coincides with the broad positive band of the blue shift. For the labeled complex, it is shifted by 50 cm$^{-1}$ from 1660~cm$^{-1}$ to 1610~cm$^{-1}$, as expected for $^{13}$C$^{15}$N-labeling\cite{Haris1992}.} \label{fig:EDIsotope}
    	\addcontentsline{toc}{subsection}{Figure~\ref{fig:EDIsotope}}
        \end{figure*} 

\newpage

        \begin{figure*} [p]
    	\centering
    	\includegraphics[clip, trim=0cm 0cm 0cm 0cm, width=0.5\textwidth]{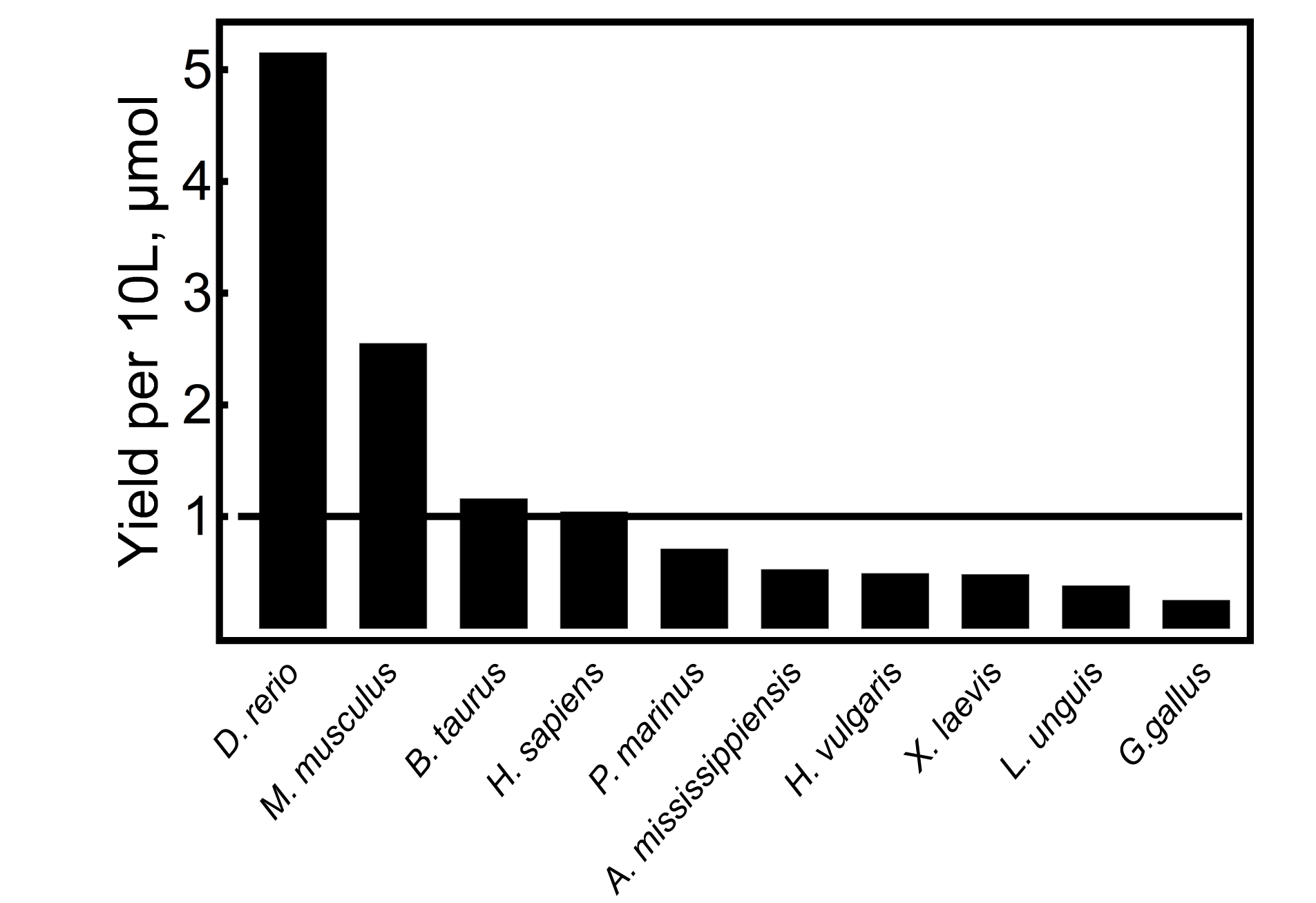}
    	\caption{Expression yield of various MCL-1 homologs from 10~L \textit{Escherichia coli} culture. A yield of 1~µmol was seen crucial for spectroscopic analysis.} \label{fig:EDYield}
    	\addcontentsline{toc}{subsection}{Figure~\ref{fig:EDYield}}
        \end{figure*}
\newpage
\clearpage
           \begin{table*}[p]
    	\centering
    	\caption{Selection of species whose MCL-1 homologs were investigated}
    	
    	\begin{tabular}{l l l l}
    		\hline
    		&MCL-1 Uniprot &divergence&confidence\\
                &ref. ID&time, Ma&interval, Ma\\
    		\hline
    		   \textit{H. sapiens} & Q07820 & 0 & -\\
                  \textit{M. musculus} & P97287 & 87 & 81-91\\
                  \textit{B. taurus}  & A5PJR2 & 94 & 92-97\\ 
                  \textit{G. gallus}  & (A0A1L1RNM6)* & 319 & 316-322\\ 
                  & XP\_046788511.1 & & \\ 
                  \textit{A. mississippiensis} & A0A151NDN7 & 319 & 316-322 \\ 
                  \textit{X. laevis} & B6V6J0 & 352 & 348-356 \\ 
                  \textit{D. rerio} & (Q1L8X3)** & 429 & 423-440\\
                  & UPI000056A44D & & \\ 
                  \textit{P. marinus} & - & 563 & 494-652\\ 
                  \textit{L. unguis} & A0A1S3IZP1\_LINUN & 708 & 627-830\\ 
                  \textit{H. vulgaris} & A1E3K7 & 715 & 604-1250 \\ 
                \hline
                \multicolumn{4}{l}{*Uniprot entry (Dec 2022) has become obsolete. NCBI reference}
                \\
                \multicolumn{4}{l}{ID is given instead. **Uniprot entry (Dec 2022) has become ob-}
                \\
                \multicolumn{4}{l}{solete. Uniparc ID is given instead. }
                \\
                \multicolumn{4}{l}{The divergence times are related to \textit{H. sapiens}.}

    	\end{tabular}
    	\label{tab:Seq_Divtimes}
    \end{table*}
        
\newpage

    \begin{table*}[p]
    	\centering
    	\caption{Time constants representing the early-, mid-, and late phase of the dynamic response.}
    	
    	\begin{tabular}{l c c c}
    		\hline
    		&$\tau_{early}$& $\tau_{mid}$&$\tau_{late}$\\
                &ns&ns&$\mu$s\\
    		\hline
    		   \textit{H. sapiens} & 1.8 & 21 & 1.8\\
                  \textit{M. musculus} & 2.7 & 31 & 1.4\\
                  \textit{B. taurus}  & 2.5 & 27 & 1.6\\ 
                  \textit{G. gallus}  & 0.9 & 28 & 1.0\\ 
                  \textit{A. mississippiensis} & 1.1 & 43 & 0.7\\ 
                  \textit{X. laevis} & 2.2 & 50 & 1.3\\ 
                  \textit{D. rerio} & 3.5 & 35 & 3.6\\
                  \textit{P. marinus} & 2.1 & 38 & 1.7\\ 
                  \textit{L. unguis} & 2.7 & 39 & 1.5\\ 
                  \textit{H. vulgaris} & 2.6 & 40 & 0.8\\
                  \textit{M. musculus} ($^{13}$C$^{15}$N) & 3.9 & 36 & 2.3\\
                \hline

    	\end{tabular}
    	\label{tab:Tau}
    \end{table*}

\end{document}